\documentclass{article}
\usepackage{amsfonts}
\usepackage{amsmath}
\usepackage{amssymb}

\input{tcilatex}

\begin{document}

\title{Spin-lattice relaxation rate of a magnetic impuritiy in the spin
degenerate Anderson model}
\author{J. W. M. Pinto and H. O. Frota \\
Departamento de F\'{\i}sica-ICE, Universidade Federal do Amazonas,\\
{69077-000, Manaus-Am, Brazil}}
\maketitle

\begin{abstract}
The renormalization group formalism was applied to calculate the
spin-lattice relaxation rate T$_{1}^{-1}$ of a well-defined magnetic moment
in the neighborhood of a spin degenerate Anderson impurity. In the Kondo
regime, T$_{1}^{-1}$ in function of $\ $the temperature $T$ presents a peak
at the Kondo temperature $T_{k}$; for $T\ll T_{k}$, the system behaves as a
heavy Fermy liquid, with an enhanced density of states, which increases with
the decreasing of the Kondo temperature; T$_{1}^{-1}T$ remains an universal
function of $T/\Gamma_{k}$ up to temperatures the order of $100\Gamma_{k}$,
where $\Gamma_{k}$ is the Kondo width; for temperatures lower then $T_{k}$,
the spin relaxation rate, T$_{1}^{-1},$ is proportional to the magnetic
susceptibility multiplied by the temperature, $\chi T$; the peak of T$%
_{1}^{-1}$ at the Kondo temperature decreases with the increasing of the
distance between the Anderson impurity and the magnetic probe.
\end{abstract}

\thanks{The authors would like to thank Ufam (the Amazonas Federal
University) and CNPq/Brazil for the financial support.}

\section{Introduction}

The one impurity spin degenerated Anderson model \cite{anderson} has been
studied by different many bodies techniques, such as Green Function,
Renormalization Group, Bethe Ansatz, Quantum Monte Carlo and combination of
Quantum Monte Carlo with the of Maximum Entropy method. The static and
dynamic properties was obtained, such as specific heat, magnetic
susceptibility, photoemission and nuclear magnetic resonance. The Electron
Spin Resonance (ESR) of a probe with a well deffined magnetic moment added
to that model remains to be better studied, which is the main point of the
present work. This model has been useful to analyze the behavior of heavy
fermions and intermediate valence compounds so that the calculation of the
ESR can contribute to a better understanding of the experimental results of
the ESR in those compounds. The ESR of a magnetic impurity embedded in a
heavy fermion compounds results in an enhancement in the Korringa rate \cite%
{krug}. However, the results for the intermediate valence compounds are
controverted. Gambke \textit{at al} \cite{gambke} associated the reduction
in the relaxation rate of Gd in CePd$_{3}$ to the intermediate valence
state. On the other hand, Heirinch and Meyer \cite{heirinch} have observed
the opposite phenomena in CeBe$_{13}$:Gd, as well Barberis \textit{at al} 
\cite{barberis} in CeIr$_{2}$:Nd and Rettori \textit{at al} \cite{rettori}
in YbInCu$_{4}$:Gd. In the Anderson model the configuration of the Anderson
ion can be changed continuously from doubly occupied to empty orbital, which
allows us to analyze the effect of the Anderson impurity orbital occupation
number on the ESR relaxation rate, from the doubly to the singly occupied
orbital regimes, passing by the valence fluctuation regime.

In the present work we hve applied the numerical renormalization group (NRG)
method of Wilson \cite{wilson,krishnamurty,oliwil} to research the
relaxation rate T$_{1}^{-1}$ of a magnetic probe embedded in a host
represented by the single impurity spin degenerate Anderson model.
Originally Wilson discretized the conduction band defining the sequence $%
\varepsilon_{j}=D\Lambda^{-j}$ (where $D$ is the half-width of the
conduction band, $\Lambda>1$ and $j=1,2,3...$) as a discrete set of
conduction states. The Anderson Hamiltonian is projected onto the basis
formed by this set and the Anderson ion states and is diagonalized
iteratively and the thermodynamic average over those discrete eigenvalues
are smooth functions of temperature. The magnetic probe relaxation rate T$%
_{1}^{-1}$, however, in a consequence of the Fermi golden rule energy
conservation, shows only discrete line transitions. To overcome this
difficulty we discretize the conduction band by the sequence $\varepsilon
_{o}=D$, $\varepsilon_{j}=D\Lambda^{-j-z}$, where $z$ is a continuous
parameter fixed arbitrarily inside intervals ($z^{\ast}$, $z^{\ast}+1$),
with $z^{\ast}\in(0,1)$, as was done in references \cite{frota} and \cite%
{makoto}. The Anderson Hamiltonian is projected onto the basis formed by the
new set and the impurity states, and is diagonalized iteratively. Continuum
spectra are obtained by averaging numerically the T$_{1}^{-1}$ over $z$, at
each temperature, in the range $z^{\ast}<z<z^{\ast}+1.$

The present paper is organized as follows: Section II details the model and
describes the relaxation rate; Section III develops the formalism to obtain
the relaxation rate from the Fermi golden rule; Section IV shows the results
and discussion of the relaxation rate, and Section V contains the
conclusions.

\section{The Model}

The model is represented by the single impurity Anderson model\ $H$ and a
magnetic probe sitting at $\overrightarrow{R_{p}}$, interacting with the
Anderson ion, sitting at $\overrightarrow{R_{f}}$, via the RKKY interaction $%
H_{x}$ \cite{rkky}. The single impurity Anderson model consists of the
kinetic energy of the free electrons ($H_{cond}$), the bound energy of the
Anderson ion orbital ($H_{orb}$) and the hybridization of this orbital with
the conduction band ($H_{hib}$),

\begin{equation}
H=H_{cond}+H_{orb}+H_{hib}.   \label{Model}
\end{equation}
The Hamiltonian of the conduction electrons is written as

\begin{equation}
H_{cond}=\dsum \limits_{\overrightarrow{k},\mu}\varepsilon_{\overrightarrow{k%
}}c_{\overrightarrow{k},\mu}^{\dagger }c_{\overrightarrow{k},\mu}, 
\label{hc}
\end{equation}
where $c_{\vec{k}\mu}^{\dagger}(c_{\vec{k}\mu})$ creates (annihilates) an
electron in the conduction band with energy $\varepsilon_{\vec{k}}$, spin $%
\mu$ and momentum $\vec{k}$, obeying the usual anticommutation relation $%
\left\{ c_{\overrightarrow{k},\mu},c_{\overrightarrow{k^{\prime}}%
,\mu^{\prime}}^{\dagger}\right\} =\delta\left( \overrightarrow {k}-%
\overrightarrow{k^{\prime}}\right) \delta_{_{\mu\mu^{\prime}}}$. The second
term in Eq. (\ref{Model}), corresponding to the Anderson ion orbital, is
given by

\begin{equation}
H_{orb}=\dsum
\limits_{\mu}\varepsilon_{f}c_{f,\mu}^{\dagger}c_{f,\mu}+Uc_{f\uparrow}^{%
\dagger }c_{f\uparrow}c_{f\downarrow}^{\dagger}c_{f\downarrow}\text{,} 
\label{hao}
\end{equation}
where the operator $c_{f\mu}^{\dagger}(c_{f\mu})$, orthonormal to the
operator $c_{\overrightarrow{k},\mu}^{\dagger}(c_{\overrightarrow{k},\mu}) $
of the conduction band, creates (annihilates) an electron in the Anderson
ion orbital ($f$) with energy $\varepsilon_{f}$ and spin $\mu$, and $U$ is
the Coulomb interaction between the electrons in that orbital. The
Hamiltonian $H_{hib}$ in Eq. (\ref{Model}) is written as

\begin{equation}
H_{hib}=V\dsum \limits_{\overrightarrow{k},\mu}\left( c_{\overrightarrow{k}%
,\mu}^{\dagger}c_{f,\mu}+c_{f,\mu}^{\dagger }c_{\overrightarrow{k}%
,\mu}\right) \text{,}   \label{hhib}
\end{equation}
where $V$ is the hybridization interaction between the electrons of the $f$
orbital and the conduction band electrons. For a moment, if we consider that
there is no coupling between the Anderson ion orbital and the conduction
electrons ($V=0$), this orbital may be empty \qquad( $n_{f}=$ $c_{f\mu
}^{\dagger}c_{f\mu}$ $=0$ and energy $0$), singly ($n_{f}=1$ and energy $%
\epsilon_{f}$) or doubly ($n_{f}=2$ and energy $2\epsilon_{f}+U$) occupied
orbital. The introduction of the interaction $V\neq0$ allows transitions
between those configurations at a rate $\Gamma=\pi\rho V^{2}$, via the
conduction band, whose density of state is $\rho$. In this case the model
can represent the intermediate valence ($\left| \Delta\right| <\Gamma$) and
the Kondo ( $\Delta,$ $-\epsilon_{f}\gg\Gamma$) regimes, where $\Delta
=\epsilon_{f}+U$ is the interconfigurational energy.

The interaction between the magnetic probe and the Anderson ion orbital ($%
H_{x}$), that will be treated as a perturbation, is given by

\begin{equation}
H_{x}=-A\left[ \Psi_{\uparrow}^{\dagger}\left( \overrightarrow{R_{p}}\right)
\Psi_{\downarrow}\left( \overrightarrow{R_{p}}\right) \mathbf{I}%
_{-}+\Psi_{\downarrow}^{\dagger}\left( \overrightarrow{R_{p}}\right)
\Psi_{\uparrow}\left( \overrightarrow{R_{p}}\right) \mathbf{I}_{+}\right] 
\text{,}   \label{hx}
\end{equation}
where $A$ is the coupling constant between the magnetic probe and the
conduction electrons, and the field operator $\Psi_{\uparrow}^{\dagger}%
\left( \overrightarrow{R_{p}}\right) $ is written as \ 
\begin{equation}
\Psi_{\mu}\left( \overrightarrow{R_{p}}\right) =\dsum \limits_{%
\overrightarrow{k}}\limfunc{e}\nolimits^{i\overrightarrow{k}\cdot%
\overrightarrow{R_{p}}}c_{\overrightarrow{k},\mu}\text{,}   \label{phi}
\end{equation}
which annihilates an electron in the Wannier state at the site of the
magnetic probe ($\overrightarrow{R_{p}}$), and $\mathbf{I}_{-(+)}$ is the
lowering (raising) spin operator of the magnetic probe ion.

Before the application of the Numerical Renormalization Group (NRG) to the
Hamiltonian given by Eq. (\ref{Model}), we will introduce a basis given by
two sets of s-wave operators, one sited at the magnetic probe and the other
at the Anderson ion, which represent the new states of the conduction band
on substitution of the operators $C_{\overrightarrow{k},\mu}$. In the
present work the conduction band obeys the linear dispersion relation $%
\varepsilon _{\overrightarrow{k}}=v_{F}k$, where the energies and moments
are measured in relation to the Fermi level and the units are taken so that $%
v_{F}=1$. Considering the energy as an isotropic function of $%
\overrightarrow{k}$, \textit{i.e.}, $\varepsilon_{\overrightarrow{k}%
}=\varepsilon_{\left| \overrightarrow{k}\right| }$, the new operators are
defined as

\begin{align}
c_{\varepsilon\mu} & =\frac{1}{\sqrt{\rho}}\sum_{\overrightarrow{k}}\limfunc{%
e}\nolimits^{i\overrightarrow{k}\cdot\overrightarrow{R}_{f}}\delta\left(
\varepsilon-\varepsilon_{k}\right) c_{\overrightarrow{k},\mu }\text{ } \\
d_{\varepsilon\mu} & =\frac{1}{\sqrt{\rho}}\sum_{\overrightarrow{k}}\limfunc{%
e}\nolimits^{i\overrightarrow{k}\cdot\overrightarrow{R_{p}}}\delta\left(
\varepsilon-\varepsilon_{k}\right) c_{\overrightarrow{k},\mu }\text{,}
\end{align}
where $c_{\varepsilon\mu}$ ($d_{\varepsilon\mu}$) is the operator which
annihilates an electron at the s-wave state of the Anderson ion (magnetic
probe), with energy $\varepsilon$ and spin $\mu$, sited at $\overrightarrow {%
R}_{f}$ ($\overrightarrow{R_{p}}$), obeying the\ anticommutation relations

\begin{align}
\left\{
c_{\varepsilon\mu}^{\dagger},c_{\varepsilon^{\prime}\mu^{\prime}}\right\} &
=\delta\left( \varepsilon-\varepsilon^{\prime}\right)
\delta_{\mu\mu^{\prime}}  \notag \\
\left\{
d_{\varepsilon\mu}^{\dagger},d_{\varepsilon^{\prime}\mu^{\prime}}\right\} &
=\delta\left( \varepsilon-\varepsilon^{\prime}\right)
\delta_{\mu\mu^{\prime}}\text{.}
\end{align}
These operators do not form an orthogonal basis, since

\begin{equation}
\left\{
c_{\varepsilon\mu}^{\dagger},d_{\varepsilon^{\prime}\mu^{\prime}}\right\} =%
\frac{\sin\left( k_{F}R\right) }{\left( k_{F}R\right) }\delta\left(
\varepsilon-\varepsilon^{\prime}\right) \delta_{\mu\mu^{\prime }}\text{,}
\end{equation}
where $k_{F}$ is the Fermi momentum and $R=\left| \overrightarrow{R_{f}}-%
\overrightarrow{R_{p}}\right| $ $\ $is the distance between the Anderson ion
and the magnetic probe. Using the Gram-Smidth orthogonalization process we
define a new operator $\bar{c}_{\varepsilon\mu}$,\ orthogonal to $%
c_{\varepsilon\mu}$, obeying the standard anticommutation relation,

\begin{equation}
\bar{c}_{\varepsilon\mu}=\frac{1}{\sqrt{1-W^{2}}}\left[ d_{\varepsilon\mu
}-W(R)c_{\varepsilon\mu}\right] \text{,}
\end{equation}
with $W(R)=\sin\left( k_{F}R\right) /\left( k_{F}R\right) $. To be used
later, the operators $f_{0\mu z}$ and $g_{0\mu z}$ are defined as

\begin{equation}
\begin{tabular}{llc}
$f_{0\mu z}$ & $=$ & $\dfrac{1}{\sqrt{2}}\int\limits_{-D}^{+D}d\varepsilon
\rho^{1/2}c_{\varepsilon\mu}$ \\ 
$g_{0\mu z}$ & $=$ & \multicolumn{1}{l}{$\dfrac{1}{\sqrt{2}}\int
\limits_{-D}^{+D}d\varepsilon\rho^{1/2}\bar{c}_{\varepsilon\mu}$,}%
\end{tabular}
\label{f0g0}
\end{equation}
where $\rho$ is the state density of the conduction band.

In terms of the operators $c_{\varepsilon\mu}$, $\bar{c}_{\varepsilon\mu} $, 
$f_{0\mu z}$ and $g_{0\mu z}$ the Hamiltonians $H_{hib}$, $H_{c}$ and $H_{x}$
are written as

\begin{equation}
H_{hib}=\sqrt{\dfrac{2\Gamma}{\pi\rho}}\left( f_{0z\mu}^{\dagger}c_{f\mu
}+h.c.\right) \text{,}   \label{hhibnovo}
\end{equation}

\begin{equation}
H_{cond}=\dint \limits_{-D}^{+D}d\varepsilon\;\varepsilon\left(
c_{\varepsilon\mu}^{\dagger}c_{\varepsilon \mu}+\bar{c}_{\varepsilon\mu}^{%
\dagger}\bar{c}_{\varepsilon\mu}\right) \text{,}   \label{hcnovo}
\end{equation}

\begin{equation}
\begin{tabular}{lll}
$H_{x}$ & $=$ & $[A_{1}(R)g_{0z\uparrow}^{\dagger}g_{0z%
\downarrow}+A_{2}(R)(g_{0z\uparrow}^{\dagger}f_{0z\downarrow}+f_{0z%
\uparrow}^{\dagger
}g_{0z\downarrow})+A_{3}(R)f_{0z\uparrow}^{\dagger}f_{0z\downarrow}]\mathbf{I%
}_{-}+h.c,$%
\end{tabular}
\label{hxnovo}
\end{equation}
where

\begin{equation}
\begin{array}{ccc}
A_{1}(R) & = & -2A\left( 1-W(R)^{2}\right) \\ 
A_{2}(R) & = & -2AW(R)\left( 1-W(R)^{2}\right) ^{1/2} \\ 
A_{3}(R) & = & -2AW(R)^{2}%
\end{array}%
\end{equation}

\section{The Formalism}

The conduction eigenstates of the Hamiltonian $H_{cond}$ is represented by
s-waves with two centers of symmetry, one sited at the Anderson ion ($H_{c}$%
),

\begin{equation*}
H_{c}=\dint \limits_{-D}^{+D}d\varepsilon\;\varepsilon
c_{\varepsilon\mu}^{\dagger}c_{\varepsilon\mu} 
\end{equation*}
and the other at the magnetic probe ($H_{\bar{c}}$),

\begin{equation*}
H_{\bar{c}}=\dint \limits_{-D}^{+D}d\varepsilon\;\varepsilon\bar{c}%
_{\varepsilon\mu}^{\dagger}\bar{c}_{\varepsilon\mu}, 
\end{equation*}
with $H_{cond}=H_{c}+H_{\overline{c}}.$ Following Wilson \cite{wilson}, both
the conduction bands, $H_{c}$ and $H_{\bar{c}}$, are discretized as a
function of the discretezation parameter $\Lambda>1$, and the continue
parameter $0<z\leq1$\cite{frota}. The electronic states are written as a
complete set of orthonormal functions \ in the domain $\left( -D,D\right) $.
For the extreme intervals, these function are given by

\begin{equation}
\Psi_{\ell}^{\left( \pm\right) }\left( \varepsilon\right) =\left\{ 
\begin{tabular}{ll}
$\dfrac{1}{\left[ D\left( 1-\Lambda^{-z}\right) \right] ^{1/2}}\limfunc{e}%
\nolimits^{\pm\;i\omega\ell\dfrac{\varepsilon}{D}}$ & for $\Lambda^{-z}<\pm%
\dfrac{\varepsilon}{D}<1$ \\ 
$0$ & out side,%
\end{tabular}
\right.
\end{equation}
where $\pm$ represent the positive or negative values of $\varepsilon/D$, $%
\ell$ is the Fourier harmonic index, which takes all integer values in the
interval $\left( -\infty,+\infty\right) $, and $\omega=2\pi/(1-\Lambda ^{-z})
$ is the fundamental frequency. In the internal intervals

\begin{equation}
\Psi_{m\ell}^{\left( \pm\right) }\left( \varepsilon\right) =\left\{ 
\begin{tabular}{ll}
$\dfrac{\Lambda^{\frac{m+z}{2}}}{\left[ D\left( 1-\Lambda^{-1}\right) \right]
^{1/2}}\limfunc{e}\nolimits^{\pm\;i\omega_{m}\ell \dfrac{\varepsilon}{D}}$ & 
for $\Lambda^{-\left( m+1+z\right) }<\pm \dfrac{\varepsilon}{D}%
<\Lambda^{-\left( m+z\right) }$ \\ 
$0$ & out side,%
\end{tabular}
\right.
\end{equation}
where $m=0,1,2,\ldots$ label the descretization band intervals, and the
parameter $\omega_{m}$ is the fundamental frequency of the \textit{m-\'{e}%
sime} interval, given by $\omega_{m}=2\pi\Lambda^{m+z}/(1-\Lambda^{-1})$.

The operators $c_{\varepsilon\mu}$ and $\bar{c}_{\varepsilon\mu}$, are
expanded in these basis as

\begin{equation}
\begin{tabular}{lll}
$c(\bar{c})_{\varepsilon\mu}$ & $=$ & $\dsum \limits_{\ell=-\infty}^{+\infty}%
\left[ a(\bar{a})_{\ell\mu}\Psi_{\ell}^{\left( +\right) }\left(
\varepsilon\right) +b(\bar{b})_{\ell\mu}\Psi_{\ell}^{\left( -\right) }\left(
\varepsilon\right) \right] +\vspace{0.08in}$ \\ 
&  & $\dsum \limits_{m=0}^{+\infty}\dsum \limits_{\ell=-\infty}^{+\infty}%
\left[ a(\bar{a})_{m\ell\mu}\Psi_{m\ell}^{\left( +\right) }\left(
\varepsilon\right) +b(\bar{b})_{m\ell\mu}\Psi_{m\ell}^{\left( -\right)
}\left( \varepsilon\right) \right] $,%
\end{tabular}
\notag
\end{equation}
with

\begin{equation}
\begin{tabular}{ll}
$a(\bar{a})_{\ell\mu}=\dint \limits_{-D}^{+D}d\varepsilon\left[
\Psi_{\ell}^{\left( +\right) }\left( \varepsilon \right) \right] ^{\ast}c(%
\bar{c})_{\varepsilon\mu}$; & $b(\bar{b})_{\ell\mu }=\dint
\limits_{-D}^{+D}d\varepsilon\left[ \Psi_{\ell}^{\left( -\right) }\left(
\varepsilon \right) \right] ^{\ast}c(\bar{c})_{\varepsilon\mu}\vspace{0.08in}
$ \\ 
$a(\bar{a})_{m\ell\mu}=\dint \limits_{-D}^{+D}d\varepsilon\left[
\Psi_{m\ell}^{\left( +\right) }\left( \varepsilon \right) \right] ^{\ast}c(%
\bar{c})_{\varepsilon\mu}$; & $b(\bar{b})_{m\ell \mu}=\dint
\limits_{-D}^{+D}d\varepsilon\left[ \Psi_{m\ell}^{\left( -\right) }\left(
\varepsilon \right) \right] ^{\ast}c(\bar{c})_{\varepsilon\mu}$,%
\end{tabular}%
\end{equation}
where $a_{\ell\mu}$, $b_{\ell\mu}$, $a_{m\ell\mu}$, $b_{m\ell\mu}$ and $\bar{%
a}_{\ell\mu}$, $\bar{b}_{\ell\mu}$, $\bar{a}_{m\ell\mu}$, $\bar
{b}%
_{m\ell\mu}$ form, respectively, complete sets of orthonormal operators
obeying the standard anticommutation relations

\begin{equation}
\begin{array}{ccc}
\left\{ a_{\ell\mu},a_{\ell^{\prime}\mu^{\prime}}^{\dagger}\right\} & = & 
\delta_{\ell\ell^{\prime}}\delta_{\mu\mu^{\prime}} \\ 
\left\{ a_{m\ell\mu},a_{m^{\prime}\ell^{\prime}\mu^{\prime}}^{\dagger
}\right\} & = & \delta_{mm^{\prime}}\delta_{\ell\ell^{\prime}}\delta_{\mu
\mu^{\prime}} \\ 
\left\{ \bar{a}_{\ell\mu},\bar{a}_{\ell^{\prime}\mu^{\prime}}^{\dagger
}\right\} & = & \delta_{\ell\ell^{\prime}}\delta_{\mu\mu^{\prime}} \\ 
\left\{ \bar{a}_{m\ell\mu},\bar{a}_{m^{\prime}\ell^{\prime}\mu^{\prime}}^{%
\dagger}\right\} & = & \delta_{mm^{\prime}}\delta_{\ell\ell^{\prime}}%
\delta_{\mu\mu^{\prime}}%
\end{array}%
\end{equation}

\bigskip Replacing the expression for the oparator $a_{\varepsilon\mu}$ into
the the expression\ for $H_{c}$ \ref{hcnovo} and eliminating the terms with $%
\ell\neq0$, approach that was successfully used to calculate the magnetic
susceptibility and the specific heat of the Kondo model \cite{wilson,oliwil}%
, the magnetic susceptibility \cite{krishnamurty} and the photoemission
spectroscopy \cite{frota} of the Anderson model, $H_{c}$\ and $H_{\bar{c}}$\
are written as

\begin{equation*}
\begin{array}{lll}
H_{c} & = & D\dfrac{1+\Lambda^{-z}}{2}[\left(
a_{\mu}^{\dagger}a_{\mu}-b_{\mu}^{\dagger}b_{\mu}\right) \\ 
&  & +\dsum \limits_{m=0}^{+\infty}\Lambda^{-\left( m+z\right) }\left(
a_{m\mu}^{\dagger}a_{m\mu}-b_{m\mu }^{\dagger}b_{m\mu}\right) ] \\ 
H_{\bar{c}} & = & D\dfrac{1+\Lambda^{-z}}{2}[(\bar{a}_{\mu}^{\dagger}\bar {a}%
_{\mu}-\bar{b}_{\mu}^{\dagger}\bar{b}_{\mu}) \\ 
&  & +\dsum \limits_{m=0}^{+\infty}\Lambda^{-\left( m+z\right) }\left( \bar{a%
}_{m\mu}^{\dagger}\bar{a}_{m\mu }-\bar{b}_{m\mu}^{\dagger}\bar{b}%
_{m\mu}\right) ]%
\end{array}
\end{equation*}
where $a(\bar{a})_{0\mu}\equiv a(\bar{a})_{\mu}$, $b(\bar{b})_{0\mu}\equiv b(%
\bar{b})_{\mu}$, $a(\bar{a})_{m0\mu}\equiv a(\bar{a})_{m\mu}$ and $b(\bar
{b%
})_{m0\mu}\equiv b(\bar{b})_{m\mu}$, for the conduction band electrons sited
at the Anderson ion orbital (magnetic probe). The operators $f_{0z\mu}$ and $%
g_{0z\mu}$ of Eq.(\ref{f0g0}) are now written as

\begin{equation}
\begin{tabular}{lll}
$f_{0z\mu}$ & $=$ & $\left( \dfrac{1+\Lambda^{-z}}{2}\right) ^{1/2}\left(
a_{\mu}+b_{\mu}\right) +\left( \dfrac{1+\Lambda^{-1}}{2}\right) ^{1/2}\dsum
\limits_{m=0}^{+\infty}\left( a_{m\mu}+b_{m\mu}\right) $ e\vspace{0.08in} \\ 
$g_{0z\mu}$ & $=$ & $\left( \dfrac{1+\Lambda^{-z}}{2}\right) ^{1/2}\left( 
\bar{a}_{\mu}+\bar{b}_{\mu}\right) +\left( \dfrac{1+\Lambda^{-1}}{2}\right)
^{1/2}\dsum \limits_{m=0}^{+\infty}\left( \bar{a}_{m\mu}+\bar{b}%
_{m\mu}\right) $,%
\end{tabular}%
\end{equation}
obeying the orthonormalization condition

\begin{align}
\left\{ f_{0z\mu},f_{0z\mu^{\prime}}^{\dagger}\right\} & =\delta_{\mu
\mu^{\prime}}  \notag \\
\left\{ g_{0z\mu},g_{0z\mu^{\prime}}^{\dagger}\right\} & =\delta_{\mu
\mu^{\prime}}\text{.}
\end{align}

The logarithmic discretization of the conduction band results in the
definition of a basis of operators $\left\{ a_{m\mu}\left( \bar{a}_{m\mu
}\right) ,b_{m\mu}\left( \bar{b}_{m\mu}\right) \right\} $, where $0\leq
m\leq\infty$, which is unsuitable for the numerical approach of the
conduction electrons-Anderson ion interaction for all order of the
hybridization $\Gamma $, since the Anderson ion orbital is coupled to all
conduction states $a_{m\mu}$ and $b_{m\mu}$ via the operator $f_{0z\mu}$.
Then a new basis $\left\{ f_{nz\mu},g_{nz\mu}\right\} $ will be defined,
where each operator $f(g)_{nz\mu}$ is coupled only to the operators $%
f(g)_{\left( n\pm1\right) z\mu}$ and solely the operator $f_{0z\mu}$ is
coupled to the Anderson ion orbital \cite{wilson}.

Taking an unitary transformation of the set of operators $\left\{ a(\bar
{a}%
)_{m\mu},b(\bar{b})_{m\mu}\right\} $ to the new set of orthonormal operators 
$\left\{ f(g)_{nz\mu}\right\} $, the conduction band Hamiltonians are given
by

\begin{align}
H_{c} & =D\frac{1+\Lambda^{-1}}{2}\dsum \limits_{n=0}^{\infty}\left(
\varepsilon_{n}^{z}f_{nz\mu}^{\mathbf{\dagger}}f_{(n+1)z\mu }+h.c.\right)
\label{hca} \\
H_{\bar{c}} & =D\frac{1+\Lambda^{-1}}{2}\dsum
\limits_{n=0}^{\infty}(\varepsilon_{n}^{z}g_{nz\mu}^{\mathbf{\dagger}%
}g_{(n+1)z\mu}+h.c.)
\end{align}
where the parameter $\varepsilon_{n}^{z}$ is numerically calculated. In the
limit of large $n$, $\varepsilon_{n}^{z}$ is approached by the asymptotic
expression \cite{frota}

\begin{equation}
\varepsilon_{n}^{z}\cong\Lambda^{\left( 1-z\right) -n/2}\text{,} 
\label{epsz}
\end{equation}
which for $z=1$ recovers the Wilson expression \cite{wilson}.

The series in the Hamltonian $H_{c}$ can be truncated without directly
affect the interaction between the electrons of the Anderson ion orbital and
the conduction band electrons. As the energy associated with the opertors\ $%
f_{nz\mu}$ and $g_{nz\mu}$, for large $n,$is of order of $\Lambda^{\left(
1-z\right) -n/2}$, one can choose a $n=N$ sufficient large so that this
quantity becomes very small as compared with the energy scale relevant to
the problem which is defined by the temperature,

\begin{equation}
D\frac{1+\Lambda^{-1}}{2}\Lambda^{\left( 1-z\right) -N/2}\ll k_{B}T\text{.}
\end{equation}
The truncated Hamiltonian $H_{c}$ and $H_{\overline{c}}$ are written as

\begin{equation*}
\begin{array}{ccc}
H_{c} & = & D\dfrac{1+\Lambda^{-1}}{2}\dsum
\limits_{n=0}^{N-1}\varepsilon_{n}^{z}\left( f_{nz\mu}^{\mathbf{\dagger}%
}f_{(n+1)z\mu }+h.c.\right) \\ 
H_{\bar{c}} & = & D\dfrac{1+\Lambda^{-1}}{2}\dsum
\limits_{n=0}^{N-1}\varepsilon_{n}^{z}\left( g_{nz\mu}^{\mathbf{\dagger}%
}g_{(n+1)z\mu }+h.c.\right)%
\end{array}
\end{equation*}

It is interesting to be observed that the truncation of the series in the
Hamiltonians $H_{c}$ doesn't affect the energy of the Anderson ion orbital,
since the coupling of the conduction electrons with the electrons of this
orbital is only via the operator $f_{0z\mu}$ of the basis $\{f_{nz\mu}\}.$

The Hamiltonian $H$ (Eq.(\ref{Model})), in terms of the operators of the
basis $\{f_{nz\mu},g_{nz\mu}\}$, can be written as

\begin{equation*}
\begin{array}{lll}
H & = & H_{A}+H_{\bar{c}} \\ 
H_{A} & = & D\left\{ \dfrac{1+\Lambda^{-1}}{2}\dsum
\limits_{n=0}^{N-1}\varepsilon_{n}^{z}\left( f_{nz\mu}^{\mathbf{\dagger}%
}f_{(n+1)z\mu }+h.c.\right) \right. \\ 
&  & \left. +\dfrac{\varepsilon_{f}}{D}c_{f\mu}^{\dagger}c_{f\mu}+\dfrac {U}{%
D}c_{f\uparrow}^{\dagger}c_{f\uparrow}c_{f\downarrow}^{\dagger
}c_{f\downarrow}+\sqrt{\dfrac{2\Gamma}{\pi D}}\left( f_{0z\mu}^{\dagger
}c_{f\mu}+h.c\right) \right\}%
\end{array}
\end{equation*}
where the state density of the conduction band $\rho=1/D$. For a better
interpretation of the results, we defne $H_{A}^{N}$ and $H_{\bar{c}}^{N}$\
as the Hamiltonian $H_{A}$ and $H_{\bar{c}}$ divided by $D(1+\Lambda^{-1})%
\Lambda^{-\left( N-1\right) /2}/2$, respectively, so that the smallest
energy scale associated to the operator $f_{\left( N-1\right) z\mu }^{%
\mathbf{\dagger}}f_{N)z\mu}+f_{Nz\mu}^{\mathbf{\dagger}}f_{\left( N-1\right)
z\mu}\ $\ and $g_{\left( N-1\right) z\mu}^{\mathbf{\dagger}%
}g_{N)z\mu}+g_{Nz\mu}^{\mathbf{\dagger}}g_{\left( N-1\right) z\mu}$ be on
the order of unity:

\begin{equation}
\begin{array}{lll}
H^{N} & = & H_{A}^{N}+H_{\bar{c}}^{N} \\ 
H_{A}^{N} & = & \Lambda^{-\left( N-1\right) /2}\left\{ \dsum
\limits_{n=0}^{N-1}\varepsilon_{n}^{z}\left( f_{nz\mu}^{\mathbf{\dagger}%
}f_{(n+1)z\mu}+g_{nz\mu }^{\mathbf{\dagger}}g_{(n+1)z\mu}+h.c.\right) \right.
\\ 
&  & \left. +\widetilde{\varepsilon}_{f}c_{f\mu}^{\dagger}c_{f\mu}+%
\widetilde{U}c_{f\uparrow}^{\dagger}c_{f\uparrow}c_{f\downarrow}^{\dagger
}c_{f\downarrow}+\Gamma^{1/2}\left( f_{0z\mu}^{\dagger}c_{f\mu}+h.c\right)
\right\} \\ 
H_{\bar{c}}^{N} & = & \Lambda^{-\left( N-1\right) /2}\dsum
\limits_{n=0}^{N-1}\varepsilon_{n}^{z}\left( g_{nz\mu}^{\mathbf{\dagger}%
}g_{(n+1)z\mu }+h.c.\right)%
\end{array}
\label{hn}
\end{equation}

where 
\begin{equation}
\begin{tabular}{lll}
$\widetilde{\varepsilon}_{f}$ & $=$ & $\dfrac{2}{1+\Lambda^{-1}}\dfrac{%
\varepsilon_{f}}{D}\vspace{0.08in}$ \\ 
$\widetilde{U}$ & $=$ & $\dfrac{2}{1+\Lambda^{-1}}\dfrac{U}{D}\vspace{0.08in}
$ \\ 
$\widetilde{\Gamma}$ & $=$ & $\left( \dfrac{2}{1+\Lambda^{-1}}\right) ^{2}%
\dfrac{2\Gamma}{\pi D}\vspace{0.08in}$.%
\end{tabular}%
\end{equation}

As $\{H_{A}^{N},H_{\overline{c}}^{N}\}=0$, the Hamiltonians $H_{A}^{N}$ and $%
H_{\bar{c}}^{N}$ are diagonalized separately and the eigenstates of $H^{N}$
are direct products of the eigenstates of\ $H_{A}^{N}$ and $H_{\bar{c}}^{N}$%
. The diagonalization of the Hamiltonians $H_{A}^{N}$ and $H_{\bar{c}}^{N}$
are performed iteractively in subspaces of the same charge and spin ($%
Q,S,S_{z}$). The diagonalization process of $H_{A}^{N}$ \ begins with the
diagonalization of the Anderson ion Hamiltonian

\begin{equation}
H_{A}^{-1}=\frac{\widetilde{\varepsilon}_{f}}{\Lambda}c_{f\mu}^{\dagger
}c_{f\mu}+\frac{\widetilde{U}}{\Lambda}c_{f\uparrow}^{\dagger}c_{f\uparrow
}c_{f\downarrow}^{\dagger}c_{f\downarrow}.
\end{equation}
In the second step we add the operator $f_{0z\mu}$ and diagonalize the
Hamiltonian

\begin{equation}
H_{A}^{0}=\frac{H_{A}^{-1}}{\sqrt{\Lambda}}+\frac{\widetilde{\Gamma}^{1/2}}{%
\sqrt{\Lambda}}\left( f_{0z\mu}^{\dagger}c_{f\mu}+c_{f\mu}^{\dagger
}f_{0z\mu}\right)
\end{equation}
in relation to a basis formed by the eigenstates of the Hamiltonian of the
first step ($H_{A}^{-1}$) and the states that are constructed by the
operators $f_{0z\uparrow}^{\dagger}$, $f_{0z\downarrow}^{\dagger}$ and $%
f_{0z\uparrow }^{\dagger}f_{0z\downarrow}^{\dagger}$ applied to the
eigenstates of $H_{A}^{-1}$. Following this procedure, the Hamiltonian $%
H_{A}^{N+1}$ is diagonalized in relation to a basis formed by the
eigenstates of $H_{A}^{N}$ and the states given by the operators $%
f_{(N+1)z\uparrow}^{\dagger}$, $f_{(N+1)z\downarrow}^{\dagger}$ and $%
f_{(N+1)z\uparrow}^{\dagger }f_{(N+1)z\downarrow}^{\dagger}$ applied to
those eigenstates. The same procedure of iteractive diagonalization is used
to diagonalize the Hamiltonian $H_{\bar{c}}^{N}$.

The ralaxation time T$_{1}\left( z\right) $ is written as the Fermi golden
rule \cite{winter}:

\begin{equation}
\dfrac{1}{\text{T}_{1}\left( z\right) }=\dfrac{4\pi}{h}\dsum
\limits_{I,F}P_{I}\left( z\right) \left| \left\langle I\left( z\right)
\right| H_{x}\left| F\left( z\right) \right\rangle \right| ^{2}\delta\left(
E_{I}\left( z\right) -E_{F}\left( z\right) \right) \text{,}  \notag
\end{equation}
where

\begin{equation}
P_{I}\left( z\right) =\dfrac{\limfunc{e}\nolimits^{-\beta E_{I}\left(
z\right) }}{\dsum \limits_{I}\limfunc{e}\nolimits^{-\beta E_{I}\left(
z\right) }}\text{,}  \notag
\end{equation}
is the statistical Boltzmann weight, $\beta=1/k_{B}T$, $k_{B}$ is the
Boltzmann constant, $T$ is the temperature and $\left| I\left( z\right)
\right\rangle $ and $\left| F\left( z\right) \right\rangle $ are the initial
and final many-particles states of the Hamiltonian $H$, with energy $%
E_{I}\left( z\right) $ and $E_{F}\left( z\right) $, respectively. In terms
of the eigenstates of the Hamiltonian given by Eq. (\ref{hn}), the
ralaxation time T$_{1}\left( z\right) $ is written as

\begin{equation}
\begin{tabular}{lll}
$\dfrac{1}{\text{T}_{1}\left( z\right) }$ & $=$ & $\dfrac{4\pi}{h}D\left( 
\dfrac{1+\Lambda^{-1}}{2}\right) \Lambda^{-\left( N-1\right) /2}\times$ \\ 
&  & $\dfrac{\dsum \limits_{I,F}\limfunc{e}\nolimits^{-\beta_{N}\Lambda^{%
\left( z-1\right) }E_{IN}\left( z\right) }\left| \left\langle I\left(
z\right) \right| H_{x}^{N}\left| F\left( z\right) \right\rangle \right|
^{2}\delta\left( E_{IN}\left( z\right) -E_{FN}\left( z\right) \right) }{%
\dsum \limits_{I}\limfunc{e}\nolimits^{-\beta_{N}\Lambda^{\left( z-1\right)
}E_{IN}\left( z\right) }}\text{,}$%
\end{tabular}
\label{rate}
\end{equation}
where 
\begin{equation*}
\beta_{N}=\left( \frac{1+\Lambda^{-1}}{2}\right) \Lambda^{-\left( N-1\right)
/2}\frac{D}{k_{B}T}\Lambda^{\left( 1-z\right) }\text{.} 
\end{equation*}
In the iteractive process we take $\beta$ as a constant, $\beta_{N}=\bar {%
\beta}$, with $\bar{\beta}$ $<1$, and $T$ as a function of $N$ ($T_{N}$),
according to Wilson method to calcuate the magnetic susceptibility of the
Kondo model \cite{wilson}, 
\begin{equation}
k_{B}T_{N}\left( z\right) =\left( \frac{1+\Lambda^{-1}}{2}\right)
\Lambda^{-\left( N-1\right) /2}\frac{D}{\bar{\beta}}\Lambda^{\left(
1-z\right) }\text{.}  \notag
\end{equation}

As a consequence of the logarithmic discretization of the conduction band,
Eq.(\ref{rate}) gives discrete lines centered at the energy $E_{IN}\left(
z\right) -E_{FN}\left( z\right) $. To transform these discrete lines in a
continuum spectrum the Dirac delta function propriety

\begin{equation*}
\delta\left( E_{IN}\left( z\right) -E_{FN}\left( z\right) \right) =\dsum
\limits_{z_{0}} \dfrac{\delta\left( z-z_{0}\right) }{\left| \dfrac{d}{dz}%
\left( E_{IN}\left( z\right) -E_{FN}\left( z\right) \right) \right|
_{z=z_{0}}}\text{,} 
\end{equation*}
was used where $z_{0}$ is the roots of the function $f\left( z_{0}\right)
=E_{IN}\left( z_{0}\right) -E_{FN}\left( z_{0}\right) $, and average $\dfrac{%
1}{\text{T}_{1}\left( z\right) }$ over $z$ inside the interval ($z^{\ast}$,$%
z^{\ast}+1$), with $z^{\ast}\in$($0,1$),

\begin{equation}
\begin{tabular}{ccc}
$\dfrac{1}{\text{T}_{1}}$ & $=$ & $\dfrac{4\pi}{h}k_{B}\bar{\beta}\dsum
\limits_{z_{0}}\Lambda^{\left( z_{0}-1\right) }T_{N}\left( z_{0}\right)
\dsum \limits_{I,F}\bar{P}_{I}\left( z_{0}\right) \dfrac{\left| \left\langle
I\left( z_{0}\right) \right| H_{x}^{N}\left| F\left( z_{0}\right)
\right\rangle \right| ^{2}}{\left| \dfrac{d}{dz}\left( E_{IN}\left( z\right)
-E_{FN}\left( z\right) \right) \right| _{z=z_{0}}}$,%
\end{tabular}
\label{T1cont}
\end{equation}
with 
\begin{equation*}
\bar{P}_{I}\left( z_{0}\right) = \dfrac{\limfunc{e}\nolimits^{-\bar {\beta}%
\Lambda^{\left( z_{0}-1\right) }E_{IN}\left( z_{0}\right) }}{\dsum
\limits_{I}\limfunc{e}\nolimits^{-\bar{\beta}\Lambda^{\left( z_{0}-1\right)
}E_{IN}\left( z_{0}\right) }}\text{.} 
\end{equation*}

\section{Results and discussion}

In this section we have shown the results of the relaxation time $1/$T$_{1}$
of the magnetic probe that interacts with the electrons of the Anderson ion
via a RKKY interaction given by Eq. (\ref{hxnovo}). In this discussion the
coupling constant $A$ between the magnetic probe and the conduction
electrons will be taken as $(h/(4\pi k_{B}\rho^{2})^{1/2}$.

We begin by verifying the accuracy of our numerical results, comparing the
numerical renormalization group results for $U=0$ with the analytical
solution \cite{gambke,pinto}. In Fig.(1) we present $1/($T$_{1}T)$ as a
function of the temperature $T$. For $k_{B}T\gg-\varepsilon_{f}$, $1/($T$%
_{1}T)$ $\ $tends for one, as is expected for pure metal. The transition
from the valence fluctuation to the doubly occupied orbital regime occurs
near the temperature in the order of $\Delta=-\varepsilon_{f}$, where $1/($T$%
_{1}T)$ presents a minimum due to the reduction of the density of state of
the conduction band close to the Anderson ion. For $k_{B}T\ll-\varepsilon_{f}
$, $1/($T$_{1}T)$ tends for a constant value of $0.9846$ (traced line), in
accordance wih the analytical result of $0.9803$ (full line) obtained from
the equation \cite{pinto}

\begin{equation}
\left( \dfrac{h}{4\pi\rho^{2}A^{2}k_{B}}\right) \frac{1}{T\text{ T}_{1}}%
=\left( \frac{\Delta^{2}}{\Delta^{2}+\Gamma^{2}}\right) ^{2}\text{,}  \notag
\end{equation}
with an error lower than $1\%$.

The analysis of the results will be carried out for $U\neq0$ and two regimes
of the parameter space of the Anderson model: the valence fluctuation and
the singly occupied orbital. We separate the results for $W$ $=1$ ($R=0$),
with the magnetic probe close to the Anderson ion site, and $W\neq1(R>0),$%
with the magnetic probe sited at a distance $R$ from the Anderson ion.

The valence fluctuation regime occurs for $-1<\Gamma/\Delta<1$ and $%
\varepsilon_{f}\gg\Gamma$. In this case, the configurations of doubly and
singly occupied orbital are mixed and the number of particle in the
fundamental state is in the interval $1<n_{f}<2$. In Fig. (2) we show $1/($T$%
_{1}T)$ for $\varepsilon_{f}=-0.01D$, $\Gamma=0.001D$ and two values of $%
\Delta$ ($0.0005D$ and $0.0007D$), for $W=1(R=0)$, \textit{i.e.}, with the
magnetic probe close to the Anderson ion site. For $k_{B}T\gg-\varepsilon_{f}
$,$-\Delta$ all the configurations ($n_{f}=0,1,2$) are thermically populated
and $1/($T$_{1}T)$ \ tends to one, as in pure metal. Lowering $T$, for $%
T\approx\Delta$, the Anderson ion orbital is in the valence fluctuation
regime, where the coupling between the two configurations ( $n_{f}=1$ and $%
n_{f}=2$) reduces the number of conduction states around the magnetic probe
site, with the corresponding reduction in $1/($T$_{1}T)$. For $k_{B}T\ll
\Delta$, the Anderson ion orbital is strongly coupled to the conduction
electrons and the system behaves as a Fermi sea, with $1/($T$_{1}T)\ $lower
than one. For low $T,$ $1/($T$_{1}T)$ increases with the increasing of $%
\Delta$. For high $\Delta$, the contribution of the configuration $n_{f}=1$
for the fundamental state increases and the Anderson ion orbital behaves as
a Kondo impurity, which will be discussed as follows.

The Anderson ion orbital is singly occupied when $\Delta$, $-\varepsilon
_{f}\gg\Gamma$. The interaction $U$ between the electrons inside this
orbital increases the energy of the configuration $n_{f}=2$, so that in the
fundamental state the Anderson ion orbital is in the configuration $n_{f}=1$%
. At low $T$ ($k_{B}T\ll\Gamma$) the electrons of the Anderson ion orbital
is strongly coupled to the electrons of the conduction band, with virtual
transition between the configuration $n_{f}=1$ and $n_{f}=2$, with the
additional electron coming from the conduction band. The spin states ($%
n_{f}=1,\mu=-1/2$ and $n_{f}=1,\mu=1/2$) are doubly degenerated, so that
there is an effective antiferromagnetic exchange interaction between the
electrons of the Anderson ion orbital and the conduction electrons, breaking
the degeneracy of the configurations $\mu=-1/2$ and $\mu=1/2$, arising the
well known Kondo effect. From the coupling between these configurations
emerge a singlet and a triplet state, with an energy separation $k_{B}T_{k}$%
, where $T_{k}$ is known as the Kondo temperature. A measure with
characteristic time $\gtrsim\hbar/k_{B}T_{k}$ can detect the spin inversion
of a electron of the Anderson ion orbital, which occurs in this period of
time. In order for the total magnetic moment of the fundamental state to
remain zero, the electrons of the conduction band go along with the spin
inversion of the electron of the Anderson ion orbital, so that the electrons
with energy the order of $k_{B}T$ stay a long time around these orbital,
increasing the scattering of the conduction electrons, with an increasing in
the spin relaxation rate. In Fig.(3) we present $1/$T$_{1}$ as a function of
the temperature for the parameters of the model representing the Kondo
regime, $\varepsilon_{f}=0.1D$, $\Gamma=0.011D$ and three values for $\Delta$
($0.05D$ (A), $0.07D$ (B), $0.09D$ (C)), considering $W=1(R=0)$, the
magnetic probe close to the Anderson ion site. The peaks occur around $%
\Gamma_{k}$ (the Kondo width), which is given by

\begin{equation}
\Gamma_{k}=\frac{T_{k}}{2\pi\times0.103},
\end{equation}
and takes the values $\Gamma_{k}=1.26\times10^{-4}D$ (A), $4.01\times10^{-5}D
$ (B) and $1.65\times10^{-5}D$ (C). The Kondo temperature $T_{k}$ is
calculated from the magnetic susceptibility results \cite{wilson}, using the
K. G. Wilson criterion

\begin{equation}
\frac{\chi\left( T\right) }{\left( \text{\textbf{g}}\mu_{B}\right) ^{2}}\cong%
\frac{0,10}{k_{B}T_{K}}\text{ for }T\ll T_{K}\text{,}
\end{equation}
where $\chi\left( T\right) $ is the magnetic susceptibility, \textbf{g} is
the Land\'{e} factor and $\mu_{B}$ Bohr magneton.

For low temperature, the heavy Fermi liquid behavior is better observed in
Fig. (4), where we show $1/($T$_{1}T)$ as a function of $T$, using the same
parameters of Fig.(3). For $k_{B}T\ll T_{k}$, \ $1/($T$_{1}T)$ is
drastically enhanced as compared with the normal metal results, and
increases as the Kondo temperature $T_{k}$ decreases.

The thermodynamics proprieties of the Anderson model in the Kondo regime,
like the magnetic susceptibility, are characterized by a universal behavior 
\cite{krishnamurty} when the temperature is scaled by the Kondo temperature $%
T_{k}$. This universal behavior is also observed in the electron spin
resonance when the spin relaxation rate is multiplied by the temperature.\
In Fig. (5) we show the universality of $T/$T$_{1}$ in the Kondo regime for $%
\varepsilon_{f}=0.1D$, $\Gamma=0.001D$ and three values for $\Delta$ ($0.05D$
(A), $0.07D$ (B), $0.09D$ (C)), considering $W=1(R=0)$. The universality
remains for temperatures the order of 100$\Gamma_{k}$.

In Fig. (6) we compare the spin relaxation time $1/$T$_{1}$ with the
magnetic susceptibility $\chi$ of the Anderson ion orbital. For temperatures
lower then $T_{k}$ the spin relaxation rate $1/$T$_{1}$ is proportional to $%
\chi T$, in qualitative agreement with the experimental results of Coldea 
\textit{et al}. \cite{coldea} for Kondo lattice.

Increasing the distance between the Anderson ion orbital and the magnetic
probe, $W\neq1(R>0)$, decreases the contributions of the term $%
A_{3}(R)f_{0z\mu}^{\dagger}f_{0z\mu^{,}}$ and enlarges the contributions of
the terms $A_{1}(R)g_{0z\mu}^{\dagger}g_{0z\mu^{,}}$, where $A_{1}\varpropto
(1-W(R)^{2})$, $A_{3}\varpropto W(R)^{2}$ and $W(R)=\sin(kR)/(kR)$. On the
other hand, the contribution of the term $A_{2}(R)(f_{0z\mu}^{\dagger}g_{0%
\mu^{,}}+g_{0z\mu}^{\dagger}f_{0z\mu^{,}})$, where $A_{2}\varpropto W(R)%
\sqrt{(1-W(R)^{2})}$, increases with the reduction of $W(R)$ until reaching
its maximum value $W(R)=\sqrt{2}/2$, after which it decreases until zero at
the limit $R\rightarrow\infty$. In this limit only the term $%
A_{1}(R)g_{0z\mu}^{\dagger}g_{0z\mu^{,}}$ survives , so that the spin
relaxation rate is equal to a pure metal. In Fig.(7) we present the spin
relaxation rate $1/$T$_{1}$ for $W(R)=1.0,0.7,0.5 $, and $0.3$, with the
parameter of the Anderson model given by $\varepsilon_{f}=-0.1D$, $%
\Gamma=0.011D$ and $\Delta=0.05D$. The Kondo peak, which is sited at the
temperature the order of the Kondo width $\Gamma_{k}=1.258\times10^{-4}D$,
decreases with the increasing of the distance $R$ between the Anderson ion
and the magnetic probe. The inset of the Fig.(7) shows that the Kondo effect
remains even for large distances, being eliminated only in the limit $%
R\rightarrow\infty$ ($W\rightarrow0$).

The Fermi liquid behavior is analyzed in Fig. (8), where we present $%
\gamma=1/($T$_{1}T)$ as a function of $\Delta/\Gamma$ at very low
temperature, taking $\Gamma=0.011D$, $\varepsilon_{f}=-0.10D$ and four
values $W$ ($1.0$, $0.7$, $0.5$ and $0.3$). For a fixed value of $%
\varepsilon_{f}$, the ratio $\Delta/\Gamma$ varies continuously from $%
\Delta\ll-\Gamma$ to $\Delta\gg\Gamma$, crossing three regimes: the regime
of Anderson ion orbital doubly occupied ($\Delta\ll-\Gamma$), the
intermediate valence regime ( $-\Gamma<\Delta<\Gamma$), and the Kondo regime
($\Delta\gg\Gamma$) . For $\Delta\ll-\Gamma$ the Anderson ion orbital is
doubly occupied and $\gamma$ is up bounded by one. By increasing $\Delta$
the energy of the configuration with doubly occupied orbital approaches the
energy of the configuration with a singly occupied orbital, and $\gamma$
decreases monotonically to zero, where these two configurations are
degenerated. For $\Delta\gg\Gamma$, in the Kondo regime, the single orbital
configuration becomes dominant and $\gamma$ suffers a huge increase. For $%
-\Gamma<\Delta<\Gamma$, in the intermediate valence regime, depending on the
ratio $\Delta/\Gamma$, the constant $\gamma$ can be smaller or greater than
one, so that T$_{1}^{-1}$ can be reduced or enhanced in relation to a pure
metal relaxation rate. This is in contrast with the particular case of
Coulomb interaction $U=0$ used by Gambke \textit{at al} \cite{gambke} to
analyze intermediate valence compound, in which $\gamma$ is always reduced
in relation to the Korringa relaxation rate of pure metal.

\section{Conclusion}

We have calculated the relaxation rate $1/$T$_{1}$ of a magnetic probe sited
at a distance $R$ from the impurity of the spin degenereted Anderson model,
using numerical renormalization group formalism. The numerical result is in
very good agreement with the analytical result for $U=0$. In the valence
fluctuation regime the curve $\gamma=1/(T$T$_{1})$ presents a depression at
temperature the order of $\Delta$, which is the difference between the
energy of the configurations $n_{f}=1$ and $n_{f}=2$, and behaves as a Fermi
liquid for very low temperature, augmenting $\gamma$ as $\Delta$ increases.

In the Kondo regime $1/$T$_{1}$ presents a peak at the Kondo width $\Gamma
_{k}$, which height increases as $\Gamma_{k}$ decreases. For temperature
much lower then $\Gamma_{k}$ the system behaves as a Fermi liquid,
presenting a dramatic enhancement in the relaxation rate. The product of the
temperature by the relaxation rate ($T/$T$_{1}$) as function of the
temperature scaled by $\Gamma_{k}$ obeys an universal function, even for
temperatures of the order of $100\Gamma_{k}$. For temperature lower then $%
\Gamma_{k}$, the relaxation rate $1/$T$_{1}$ is proportional to the product
of the temperature by the magnetic susceptibility ($T\chi$). Increasing the
distance $R$ between the magnetic probe and the Anderson impurity, the peak
at $\Gamma_{k}$ decreases until it disappears at $R\rightarrow\infty$. For
very low temperatures, $\gamma=1/(T$T$_{1})$ as a function of $\Delta/\Gamma$
approaches to one in the region where the Anderson impurity is doubly
occupied ($\Delta/\Gamma<1$), presents a minimum \ in the region of valence
fluctuation regime ($-1<\Delta/\Gamma<1$), and is drastically enhanced in
the region of the Kondo regime ($\Delta/\Gamma>1$).

\bigskip

\begin{center}
\newpage%

FIGURE CAPTIONS
\end{center}

Fig. 1. Numerical renormalization group result for $1/$(T$_{1}T$) as a
function of the temperature (full line) for $U=0$ and $R=0$, as compared
with the analytical result (traced line) [12], with a lower than $1\%$ error.

Fig. 2. The rate $\gamma=1/$($T$T$_{1}$) as a function of the temperature
for the valence fluctuation regime and $R=0$. For $k_{B}T\gg-\varepsilon_{f}$%
, all the Anderson ion states are thermically accessible and the system
behaves as a pure metal; for $k_{B}T\thickapprox\Delta$, $1/$($T$T$_{1}$)
has a minimum due to the reduction in the density of states around the
magnetic probe; for $k_{B}T\ll\Gamma$, $1/$($T$T$_{1}$) is constant and
increases with $\Delta$.

Fig. 3. The spin relaxation rate T$_{1}^{-1}$ as a function of the
temperature for three sets of parameters of the Anderson model, taking fixed 
$\varepsilon_{f}$ and $\Gamma$, and varying $\Delta.$ The curves present a
peak at the Kondo width $\Gamma_{k}=1.26\times10^{-4}D$ (curve A), $\Gamma
_{k}=4.01\times10^{-5}D$ (curve B) and $\Gamma_{k}=1.65\times10^{-5}D$
(curve C), which are marked in the temperature axis. $\Gamma_{k}$ was
obtained from the magnetic susceptibility, as given by Eq. (39), using the
Wilson relation $\Gamma_{k}=T_{k}/(2\pi\times0.103)$. The height of the peak
grows with the reduction of the \ Kondo temperature.

Fig. 4. The rate $\gamma=1/$($T$T$_{1}$) as a function of the temperature
for the same parameters of Fig.(3). For high $T$, the system behaves as a
pure metal; lowering the temperature, the curves present a minimum,
associated with the valence fluctuation regime; for $T\ll\Gamma_{k}$, the $%
\gamma$ is enhanced in the Kondo regime, and the system behaves as a heavy
Fermi liquid.

Fig. 5. In the Kondo regime, $T/$T$_{1}$ behaves as a universal function of
the $T/\Gamma_{k}$. The figure presents three sets of Anderson model
parameter in the Kondo regime, and the universality behavior of $T/$T$_{1}$
remains until temperatures of the order of 100$\Gamma_{k}.$

Fig. 6. In the Kondo regime, the spin relaxation rate T$_{1}^{-1}$ is
proportional to $\chi T$. The figure presents T$_{1}^{-1}$ and $\chi T$ for
the same set of parameters of the Anderson model, in the Kondo regime, where
T$_{1}^{-1}$ is proportional to $\chi T$ for temperature lower than the
Kondo temperature (marked in the temperature axis).

Fig. 7. Spin relaxation rate T$_{1}^{-1}$ as a function of the temperature
for four distances $R$ between the Anderson ion and the magnetic probe,
corresponding to $W=1$ (the magnetic probe very close to the Anderson ion),
and $W=0.7$, $0.5,$ and $0.3$. The Kondo peak decreases with the distance $%
R. $ Even for long distance (curve D), the peak survives, as is shown in the
inset.

Fig. 8. The rate $\gamma=1/$($T$T$_{1}$) as a function of $\Delta/\Gamma$,
for very low temperature and deferent distances between the Anderson ion and
the magnetic probe. For $\Delta/\Gamma\ll-1$, the Anderson ion is doubly
occupied, for $-1<\Delta/\Gamma<1$ is in the intermediate valence regime,
and for $\Delta/\Gamma\gg1$, is in the Kondo regime. The curves are lower
than one for the Anderson ion in the doubly occupied regime, present a
minimum in the valence fluctuation regime, and are enhanced in the Kondo
regime.

\end{document}